\documentclass[aps,10pt,twocolumn,amsmath,amssymb]{revtex4}
\usepackage{hyperref}
\usepackage{url}
\usepackage{graphicx}
\usepackage{epstopdf}
\usepackage[utf8x]{inputenc}
\usepackage{amssymb}
\usepackage{amsmath}

\begin{document}

\title{Soft self-assembled nanoparticles with temperature-dependent properties}
\author{Lorenzo Rovigatti}
\email{lorenzo.rovigatti@univie.ac.at}
\affiliation{Faculty of Physics, University of Vienna, Boltzmanngasse 5, A-1090 Vienna, Austria}
\author{Barbara Capone} 
\affiliation{Faculty of Physics, University of Vienna, Boltzmanngasse 5, A-1090 Vienna, Austria}
\author{Christos N. Likos} 
\affiliation{Faculty of Physics, University of Vienna, Boltzmanngasse 5, A-1090 Vienna, Austria}

\begin{abstract}
The fabrication of versatile  building blocks that are reliably self-assemble into desired ordered and disordered phases is amongst the  hottest topics in contemporary material science. To this end, microscopic units of varying complexity, aimed at assembling the target phases, have been thought, designed, investigated and built. Such a path usually requires laborious fabrication techniques, especially when a specific funcionalisation of the building blocks is required. 
Telechelic star polymers, i.e., star polymers made of a number $f$ of di-block copolymers consisting  of solvophobic and solvophilic monomers grafted on a central anchoring point, spontaneously self-assemble into soft patchy particles featuring attractive spots (patches) on the surface. 
Here we show that the tunability of such a system can be widely extended by controlling the physical and chemical parameters of the solution. Indeed, at fixed external conditions the self-assembly behaviour depends only on the number of arms and/or on the ratio of solvophobic to solvophilic monomers. However, changes in temperature and/or solvent quality makes it possible to reliably change the number and size of the attractive patches. This allows to steer the mesoscopic self-assembly behaviour without modifying the microscopic constituents. 
Interestingly, we also demonstrate that diverse combinations of the parameters can generate stars with the same number of patches but different radial and angular stiffness. This mechanism could provide a neat way of further fine-tuning the elastic properties of the supramolecular network without changing its topology.
\end{abstract}

\maketitle

\section{Introduction}

Designing novel  materials on the nanometer scale requires a careful choice of the microscopic
building blocks~\cite{bookcolloid}. In the last decade, theoretical and numerical research has demonstrated that adding 
anisotropy to selected building blocks greatly enlarges the realm of possibility~\cite{GlotzNatMat04}. 
Indeed, photonic materials, lightweight gels, self-healing plastics and devices for medical imaging 
and drug delivery have all been realised \textit{in silico}~\cite{frank_vitrimers} and, to a lesser extent, in 
experiments~\cite{granicknature,biffi_dna}. Many interesting effects arising in these systems can be
rationalised in terms of a \textit{reduced valence}: the anisotropic nature of the interaction 
limits the number of reversible bonds that each nanoparticle can establish, effectively stabilising
open, i.e. low-density, ordered and disordered structures~\cite{likosreview}. A simple yet very successful
toy model with a built-in limited valence is provided by the so-called \textit{patchy particles}, e.g., 
colloids decorated with attractive spots (patches) on their surface~\cite{Manoh_03,dayang_patchy,Zhang_04,likosreview}.
The richness of their phase behaviour, ranging from low-density reversible gels~\cite{bian} to open 
crystals~\cite{doye_tetra,granicknature,flavio_nature_comm} and cluster phases~\cite{janusprl,Rovigatti2013b}, 
spurred the development of new methods for their synthesis~\cite{Kretz_2008,wang2012colloids,sacanna_review}.
However, the fabrication of bulk quantities of monodisperse patchy colloids with tunable interactions 
has been not achieved yet.
Recently, the idea of using polymer-based systems to synthesise anisotropically-interacting particles has been 
proposed~\cite{soft_nano_patchy_nl,telechelics_letter}. In particular, a very promising idea revolves around 
telechelic star polymers (TSP), which can be
already readily synthesised, for example by using polybutadiene stars functionalised with zwitterionic end 
groups~\cite{pitsikalis_tsp,vlassopoulos_tsp_1999,vlassopoulos_tsp_2000}. 
TPS's are macromolecules made of a number $f$ of diblock copolymers grafted on a central anchoring 
point~\cite{loverso:prl:2006,loverso:jpcc:2007,likos_telechelic,telechelics_letter,koch:sm:2013}. Each of the $f$ 
diblock-co-polymeric arms is made of a ratio of $\alpha$ solvophobic  and $(1-\alpha)$ solvophilic monomers; the 
dual nature of their arms makes TSP's particularly sensitive to variations of the external conditions, such as 
temperature or ionic strength for the case of zwitterionic telechelics, and allows each particle to self-assemble 
into a soft particle with attractive patches on the surface. As a result, TSP's undergo a hierarchical self-assembly: 
on the single-scale 
particles can be tuned to self-assemble into building blocks with predetermined properties; on a larger scale, 
particles can then self-assemble into meso- and macroscopic structures~\cite{telechelics_letter,capone:njp:2013} 
which can be exploited in material science and for medical applications~\cite{helms2006dendrimers,SakaiKato201576}.

Polymeric molecular building blocks present several advantages. In terms of synthesis, no cumbersome preparation techniques are 
required~\cite{Alward, Alward2}. From a theoretical point of view, the self-assembled nature of the particles makes them 
inherently soft and floppy, providing additional control over the target structure and its properties~\cite{Rovigatti2014a,soft_patchy_bianchi}.
However, such a subtle dependence of the bulk properties on the single-star conformation calls for a  precise determination of the latter. In this work, we carry out extensive simulations of single, large telechelic star polymers for a wide range of parameters, characterising the self-assembly process  and the resulting conformation as functionality, diblock copolymer length and solvent quality vary.
We show that by tuning the chemical and physical  parameters in solution, it is possible to influence and control the number and size of the attractive patches that each particle  forms. Interestingly, we also demonstrate that different combinations of the parameters can generate stars with the same number of patches but  different radial and angular stiffness. This mechanism could provide a neat way of tuning the elastic properties of the supramolecular network without changing its topology.

\section{Model and methods}
\label{sec:methods}

We simulate TSP's made of $f$ diblock copolymer chains, anchored to a central point through their athermal parts. Each chain is made of $N_A$ monomers 
of type $A$ (solvophilic) and $N_B$ monomers of type $B$ (solvophobic). We define the fraction  of monomers of type $B$ as $\alpha = N_B / (N_A + N_B)$. 
We fix the minimum number of $A$ and $B$ monomers per chain to $N_A^{\rm min} = 40$ and $N_B^{\rm min} = 80$, respectively.
The resulting stars are thus comparable with experimental systems~\cite{pitsikalis_tsp,vlassopoulos_tsp_1999,gauthier_few_monomers}.
Bonded neighbours, i.e. particles which share a backbone link, are kept close together by a FENE potential of the form

\begin{equation}
\label{eq:fene}
V_{\rm F}(r) = -15 \epsilon \frac{r_{\rm F}^2}{\sigma^2} \log\left(1 - \frac{r^2}{r_{\rm F}^2}\right)
\end{equation}
\noindent
where $r_{\rm F}$ is the allowed maximum distance between monomers. We set $r_{\rm F}=1.5\sigma$. $\epsilon$ 
is the interaction strength. In what follows, we set $\sigma = 1$, 
$\epsilon = 1$ and also $k_{\rm B} = 1$ (Boltzmann's constant) and we express all dimensional quantities 
(length, density and temperature) in these units.

All the repulsive interactions acting between both bonded and non-bonded $A-A$ and $A-B$ pairs are modelled 
through a generalised Lennard-Jones (LJ) potential,

\begin{equation}
\label{eq:repulsive_LJ}
V_{AA}(r) = V_{AB}(r) = \left\{ \begin{array}{ll}
4\epsilon \left[ \left(\frac{\sigma}{r}\right)^{48} - \left(\frac{\sigma}{r}\right)^{24} \right] + \epsilon & \mathrm{if}\; r < r_{\rm rep}^c,\\
0 & \mathrm{otherwise}
\end{array}
\right.
\end{equation}
\noindent
with $r_{\rm rep}^c = 2^{\frac{1}{24}}\sigma \approx 1.03$. 
Finally, the attraction between the terminal solvophobic monomers is provided by the attractive tail of the same 
generalised LJ potential as in Eq.~\ref{eq:repulsive_LJ}, rescaled by a parameter $\lambda$:

\begin{equation}
\label{eq:attractive_LJ}
V_{BB}(r) = \left\{ \begin{array}{ll}
V_{AA}(r) - \epsilon\lambda & \mathrm{if}\; r < r_{\rm rep}^c,\\
4\epsilon\lambda \left[ \left(\frac{\sigma}{r}\right)^{48} - \left(\frac{\sigma}{r}\right)^{24} \right] & \mathrm{otherwise}
\end{array}
\right.
\end{equation}
\noindent
Therefore, the $\lambda$ parameter plays the role of an inverse temperature for the $B-B$ interaction. The value of 
$\lambda$ at which purely solvophobic chains have a Gaussian statistics, equivalent to the so-called $\theta$-temperature,
is $\lambda_\theta \approx 0.92$.
For performance reasons we truncate and shift this potential at $r_c = 1.5$.

We run Brownian Dynamics simulations at fixed temperature $kT/\epsilon = 0.5$~\cite{John}. 
Single TSP's with functionality $f$ ranging between $3$ and $18$  and values of $\alpha$ ranging between $0.3$ and $0.8$ are investigated.
In this work we characterise how chemical and physical parameters can influence single star properties, self-assembling behaviour, localisation and flexibility of the patches, both angular and radial, focussing on monomer-resolved stars so as to access a broad temperature range and investigate a large number of $(f,\alpha)$ combinations.

Recent studies \cite{soft_patchy_bianchi} showed that soft patchy particles assemble into different gel-like structures depending on the softness of both angular and radial position of the patches with respect to the equilibrium position; at the same time, works on coarse-grained telechelic star polymers~\cite{capone:njp:2013}  showed that the single star self-aggregating scenario is preserved upon increasing density in solution, for stars with various different $(f,\alpha)$ combinations. 
It hence becomes important to completely characterise, on the full monomer scale, how a change of chemical (solvent quality e.g. $\lambda^{-1}-$temperature effect), and physical parameters (such as $(f,\alpha)$ combinations) can lead to the formation of particles with a given number of patches, and how the radial and angular flexibility of those functionalised domains can be tuned and controlled by parameters external to the macromolecules. 
We hence carry out an extensive characterisation of the stars and of their self-assembling behaviour as a function of $\lambda, f$ and $\alpha$.

The first parameter that we use to classify the stars is the number of patches $N_p$ that the macromolecules self-assemble, defined as the number of clusters formed by multiple arms. If the interaction energy between at least two monomers of different arms is negative, i.e., if they experience a net attraction, then the two arms belong to the same
cluster, and hence to the same patch. According to this definition,  $N_p \approx 0$ in the good solvent limit ($\lambda \to 0$), since the attractive nature of the entropic-solvophilic monomers does not play any significant role in the self-aggregating behaviour that is instead driven by the enthalpic-solvophobic part of the molecule. 

We start off by making a characterisation of the stars based on the number of self-assembled functionalised regions. We then move deeper into the description of the soft molecular building blocks by quantifying how the patch population $s_p$, defined as the number of arms that form a patch, is influenced by the choice of the parameters. Stars with different compositions can assemble into soft-patchy nano building blocks decorated by the same number of functionalised regions. Their radial and angular flexibility will crucially depend on the number of arms that are participating to the formation of a patch and on the size of the patch itself. Hence we perform a radial-angular flexibility analysis by  characterising the geometry of the assembled TSP. We compute the average distance between the centre of mass of a patch and the position of the anchoring point, $r_p$, and the average angle between two patches, $\theta_p$, defined as the angle between two vectors pointing towards each pair of patches, starting from the anchoring point. 
The quantities $r_p$ and $\theta_p$ are two very important parameters to play with when looking to hierarchically self-assemble specific structures. For example, particles with an excess of radial and angular flexibility might lose the capability to crystallise~\cite{frank_nat_phys}.

The overall conformation and shape of the stars is another key characteristic, and we will elucidate its dependence on $f$, $\alpha$ and $\lambda$ .
The latter analysis is done by computing the shape anisotropy $\delta$, the prolateness $S$ and the acylindricity 
$c$~\cite{theodorou_shape,steinhauser_shape,zifferer_shape}. These quantities are derived from the gyration tensor:

\begin{equation}
G_{mn} \equiv \frac{1}{N} \sum_{i=1}^N (r_i^m - r_{\rm cm}^m) \cdot (r_i^n - r_{\rm cm}^n)
\end{equation}
where $N$ is the total number of monomers, $r_i^m$ is the $m$-th component of the position of the $i$-th monomer and 
$r_{\rm cm}^m$ is the $m$-th component of the position of the star centre of mass. Diagonalising the tensor $\textbf{G}$
yields three eigenvalues $\lambda_i$, $i=1,2,3$, which are ordered as $\lambda_1 \geq \lambda_2 \geq \lambda_3$. We use
these values to compute the aforementioned shape parameters, which are defined as follows:

\begin{eqnarray}
\delta & = & 1 - 3 \left\langle \frac{I_2}{I_1^2} \right\rangle \label{eq:def_delta}\\
S & = & \left\langle \frac{(3\lambda_1 - I_1)(3\lambda_2 - I_1)(3\lambda_3 - I_1)}{I_1^3} \right\rangle\label{eq:def_S}\\
c & = & \left\langle \frac{\lambda_2 - \lambda_3}{I_1} \right\rangle\label{eq:def_c}
\end{eqnarray}
where $I_1 = \lambda_1 + \lambda_2 + \lambda_3$ and $I_2 = \lambda_1\lambda_2 + \lambda_2\lambda_3 + \lambda_3\lambda_1$
and the angular brackets have the meaning of ensemble averages. The first parameter, $\delta$, is positive definite 
and quantifies the asphericity. $S$, which takes values between $-0.25$ and $2$, measures prolateness ($S > 0$) or 
oblateness ($S < 0$). We notice that, in the system under investigation, $\delta$ and $S$ turn out to follow the exact
same trends. Therefore, for the sake of clarity and conciseness we decided to only show the latter quantity. 
The last parameter, $c$, is always equal to or larger than $0$ and quantifies the cylindrical symmetry of the star, 
taking the value $0$ only for perfectly cylindrical conformations. It is defined as to take into account the fact that, 
as demonstrated in Section~\ref{subsec:shape}, all investigated conformations are always prolate, i.e. $S > 0$.

Finally, here and in what follows we use the expression \textit{soft particle} to refer to building blocks that are partially 
or completely interpenetrable and exhibit an intrinsic floppiness, in contrast to usual ``hard'' colloids. An overview of 
wide classes of soft particles can be found in Ref.~\cite{likos_soft_soft}. 



\section{Results}

Due to their intrinsic nature, telechelic star polymers exhibit a self-assembling behaviour arising from the competition 
between the entropic self-avoiding repulsion of the inner part (good solvent) and the enthalpic attractions amongst the 
solvophobic tails of the $f$ arms that constitute the macromolecules. It hence appears evident how a change in solvent 
quality (chemical perturbation to the system that can be performed by a change in temperature), modifies the enthalpic 
contribution, therefore affecting the whole single-macromolecule self-assembling process. 
For small values of the coupling constant $\lambda$, stars with a small percentage of attractive monomers do not have 
enough enthalpic contribution to assemble into a patchy structure. However, as soon as a minimum amount of attractive 
monomers is reached (a number that depends on the solvent quality and it is thus linked to the $\lambda$ parameter), 
patchy structures arise. 

\begin{figure*}
\begin{center}
\includegraphics[width=0.3\textwidth]{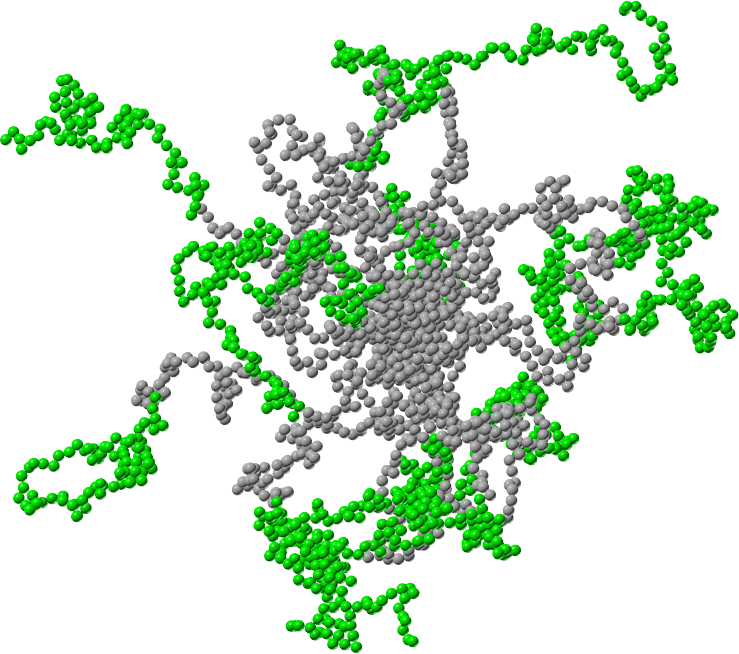}\hspace{0.1cm}
\includegraphics[width=0.26\textwidth]{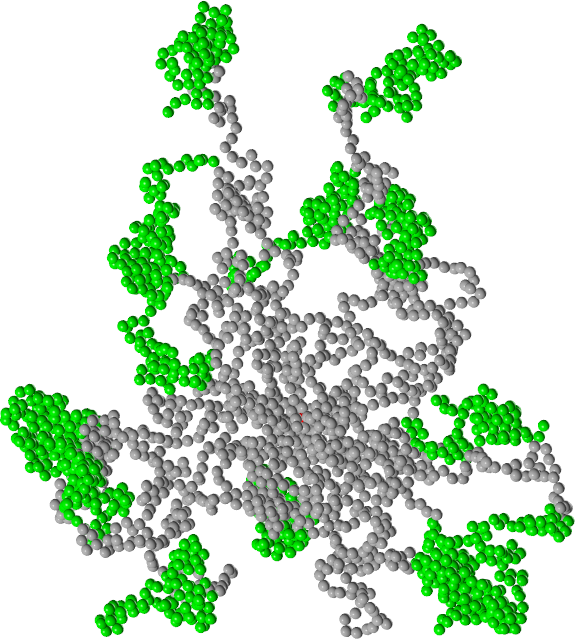}\hspace{0.3cm}
\includegraphics[width=0.19\textwidth]{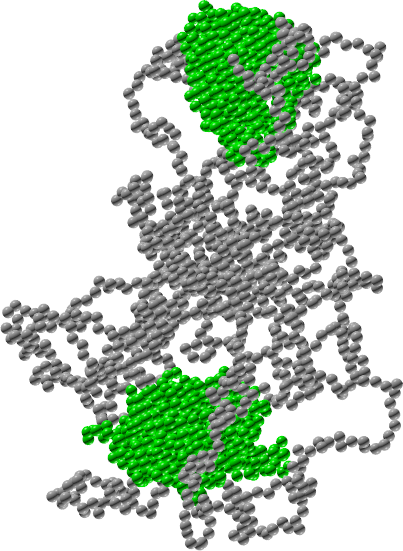}
\end{center}
\caption{\label{fig:snapshots}A TSP with $f=15$ and $\alpha=0.5$ for, from left to right, $\lambda=0.80$, $1.00$ 
and $1.10$.}
\end{figure*}

Figure~\ref{fig:snapshots} shows representative snapshots of a star with $f=15$ and $\alpha=0.5$ for  different
values of the attraction coupling constant $\lambda$. The picture sketches the self-assembly process that takes
place as $\lambda$ increases for a fixed $(f,\alpha)$ combination. When the $B$-monomers, coloured in green, are 
in a good solvent (i.e. for small values of $\lambda$), the star is open and the inter-chain attraction is 
negligible. In this regime stars resemble the usual athermal star polymers~\cite{grest_star_polymers}. Upon 
worsening the solvent-quality, the solvophobic monomers start to collapse on themselves forming patches. A further 
increase of the attraction leads to a coarsening of the patches, which decrease in number but grow in size, as 
shown in the rightmost snapshot of Figure~\ref{fig:snapshots}. Similar figures are used in the plots throughout 
the paper to increase readability and to show how stars with different parameters look like.

We note that the functionalities investigated here yield small numbers of patches, ranging from one to four. We 
will put particular emphasis on stars that exhibit one to three patches since these can be used to generate low-density
disordered (gel) phases~\cite{bian,prl-lisbona,Rovigatti2013b}. However, other combinations of $(f,\alpha)$ 
can be used to select higher-valency particles that can be used to assemble denser, and possibly ordered,
phases~\cite{telechelics_letter,capone:njp:2013}. Additional control could be provided by confining the 
system, effectively reducing its dimensionality to generate two-dimensional or quasi two-dimensional 
phases with distinct symmetries and properties~\cite{alexander_confinement,christos_confinement}.

\subsection{Characterisation of the patches}

Extensive studies of toy models of rigid hard patchy particles have shown that the single most important parameter in determining 
the overall phase behaviour of the system is the number of patches~\cite{lungo,FoffiKern}. These models usually
employ particles with fixed numbers of patches, even though it is possible to enforce a temperature-dependent
valence by using particles decorated with dissimilar patches~\cite{prl-lisbona,Rovigatti2013b}. By contrast, soft self-assembling 
systems as soft patchy particles or molecular telechelic star polymers present a variable number of patches that depends on external parameters such 
as solvent quality or temperature, role which is here played by $\lambda$. Therefore, understanding how a change in 
$\lambda$ affects the average number of patches $N_p$ for stars with fixed combinations of $f$ and $\alpha$ will allow to change 
the functionalisation  of the molecular building blocks, and hence their hierarchical self-assembling process, without the need to 
change the molecules in solution.

\begin{figure}[h!]

\includegraphics[width=0.4\textwidth]{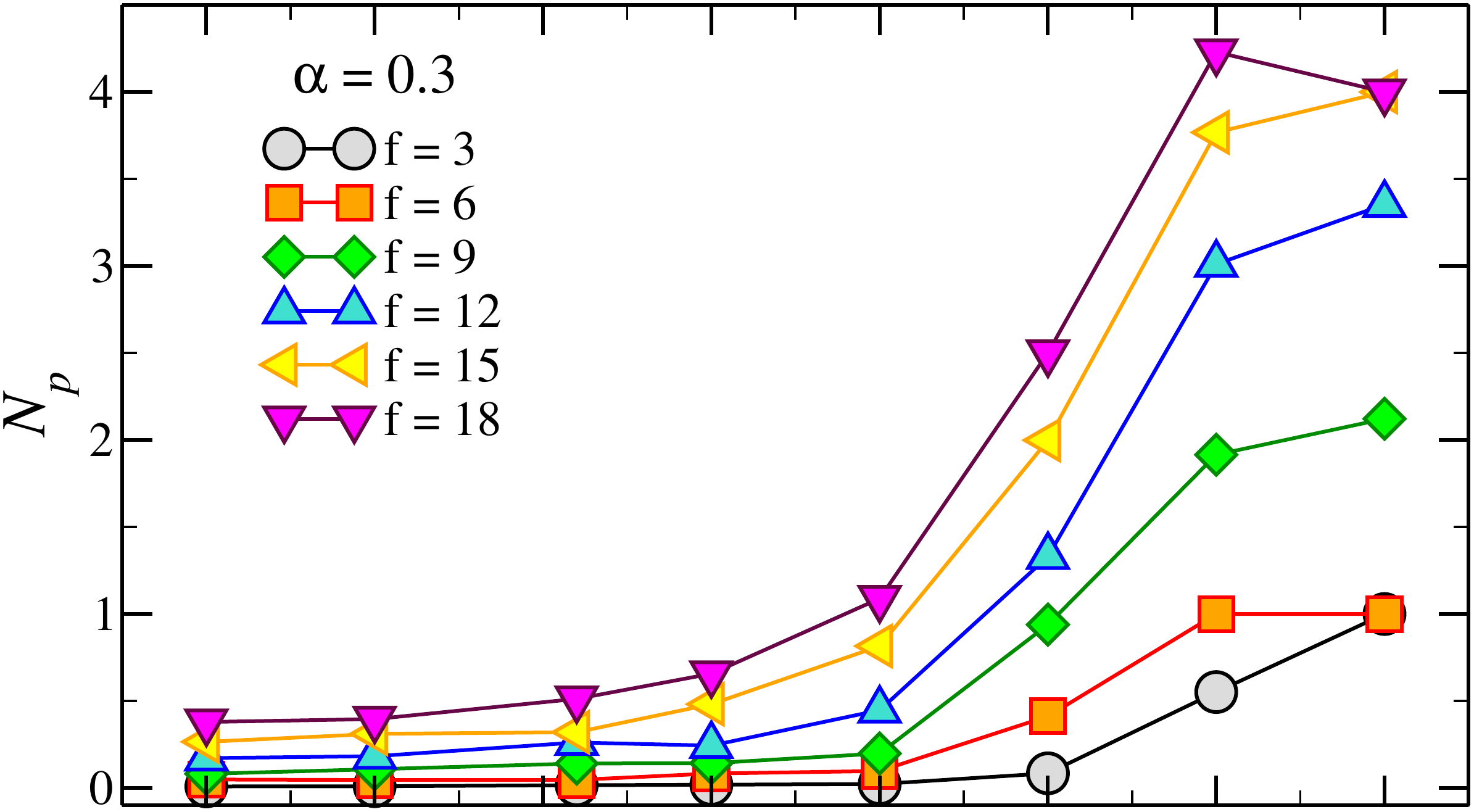}\\
\includegraphics[width=0.4\textwidth]{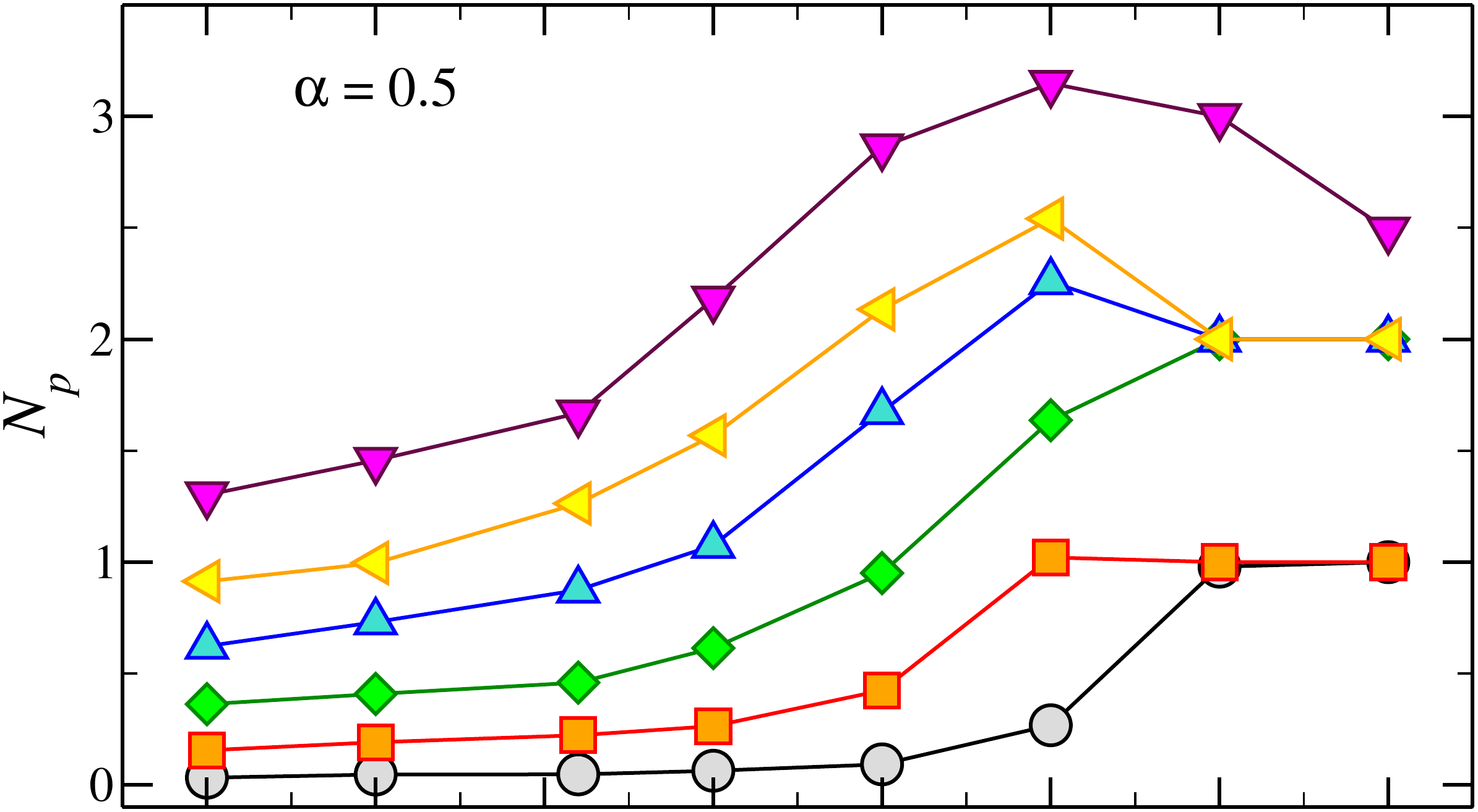}\\
\includegraphics[width=0.4\textwidth]{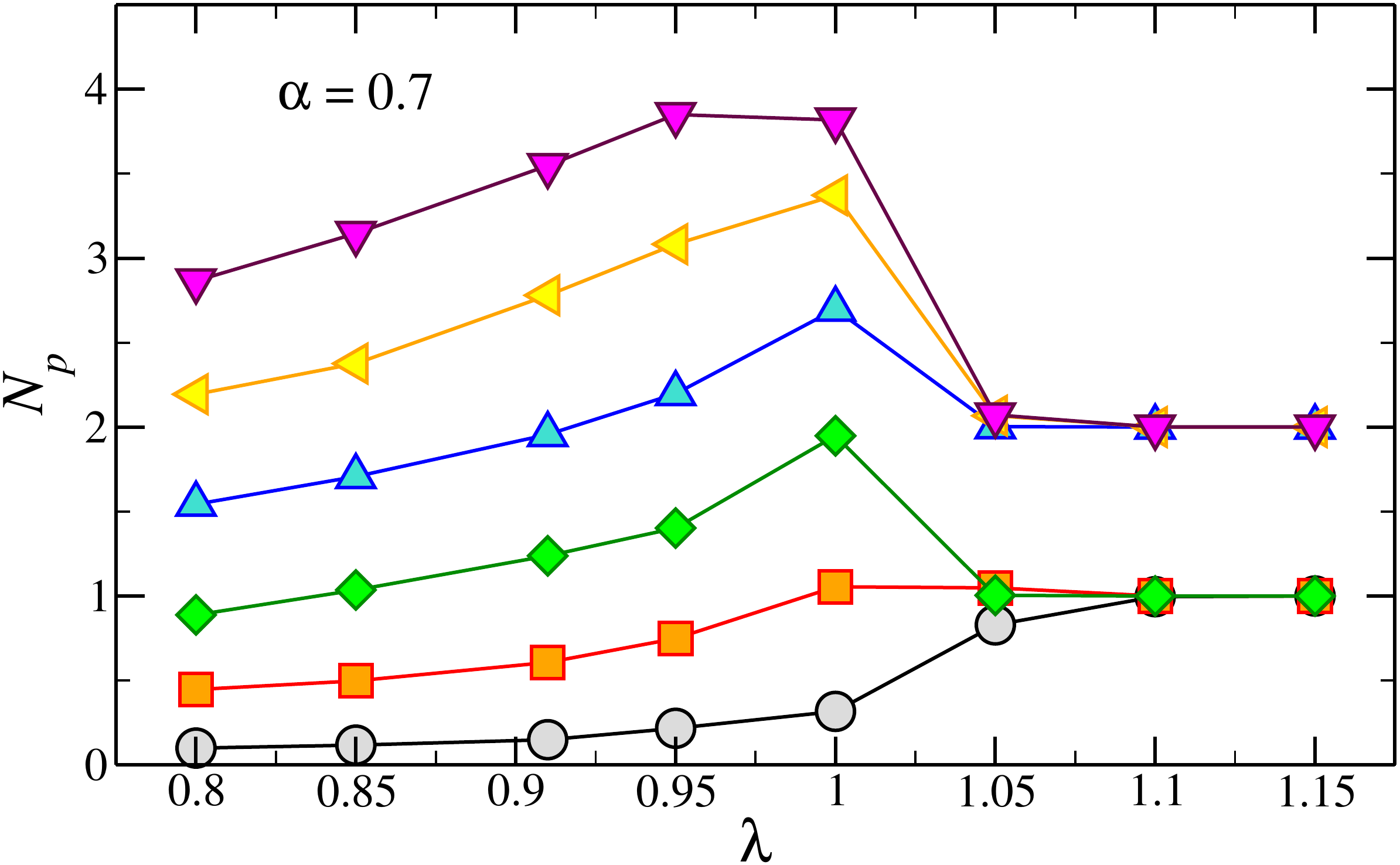}

\caption{\label{fig:alpha}Number of patches $N_p$ as a function of $\lambda$ for (top) $\alpha=0.3$, (middle)
$\alpha=0.5$ and (bottom) $\alpha=0.7$.}
\end{figure}

Figure~\ref{fig:alpha} shows $N_p$ as a function of $\lambda$ for all investigated $f$ and $\alpha = 0.3$, 
$0.5$ and $0.7$. For the lowest values of $\lambda$ investigated here, all the curves are increasing functions 
of the coupling constant, signalling the onset of the self-assembly process. A comparison between different $\alpha$ suggests that 
this onset occurs at lower values of $\lambda$ as $\alpha$ increases. 
For the lowest value of $\alpha$ considered here, all the curves but the $f=18$ one are monotonic with both $\lambda$ and $f$. 
However, for high values of $\alpha$ and $\lambda$ the curves exhibit a clear non-monotonicity. Indeed, when 
the attraction between solvophobic monomers exceeds a certain threshold, the arms start feeling a strong 
mutual attraction, collapsing on themselves and forming fewer, although larger, patches. Upon further increasing 
$\lambda$ ($\lambda > 1.15$) these high-$\alpha$ systems fall off of equilibrium and $N_p$ eventually plateaus. 
A visual inspection of the configurations shows that monomers in the largest patches eventually crystallise.

Depending on the number and size of the patches, these low-valence TSP's will assemble into different large-scale 
structures. A single patch can yield micelles or interconnected (wormlike) micelles, depending on the patch 
size~\cite{likos_telechelic}. As the number of patches increases so does the connectivity, meaning that inter-star 
bonds become more common, eventually leading to network formation. The overall properties of this network will
depend not only on the number and size of the patches, as it is the case for patchy colloids~\cite{likosreview}, 
but also on the radial and angular stiffness of the stars themselves~\cite{soft_patchy_bianchi}.

\begin{figure}[h!]

\includegraphics[width=0.4\textwidth]{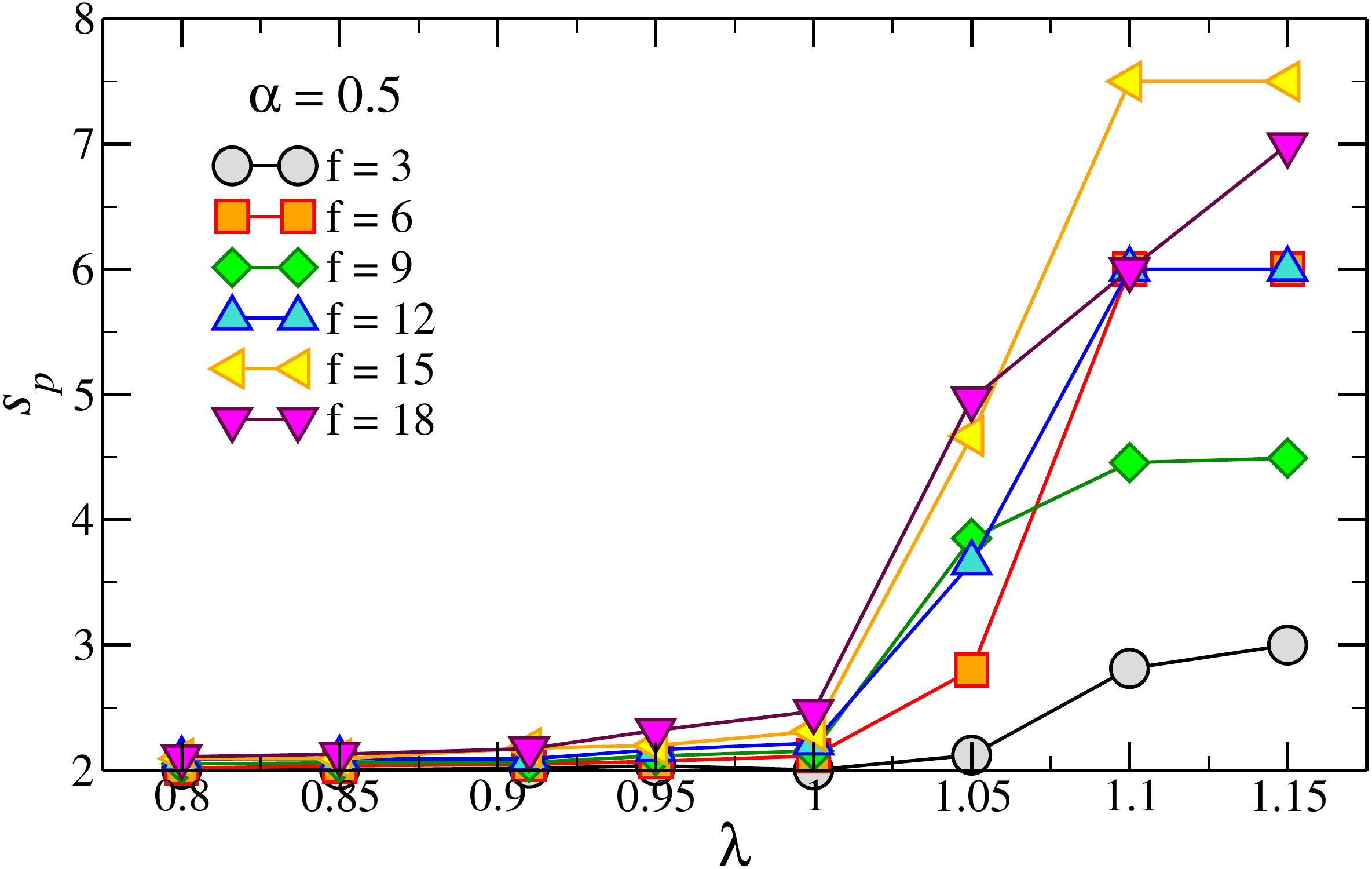}
\includegraphics[width=0.4\textwidth]{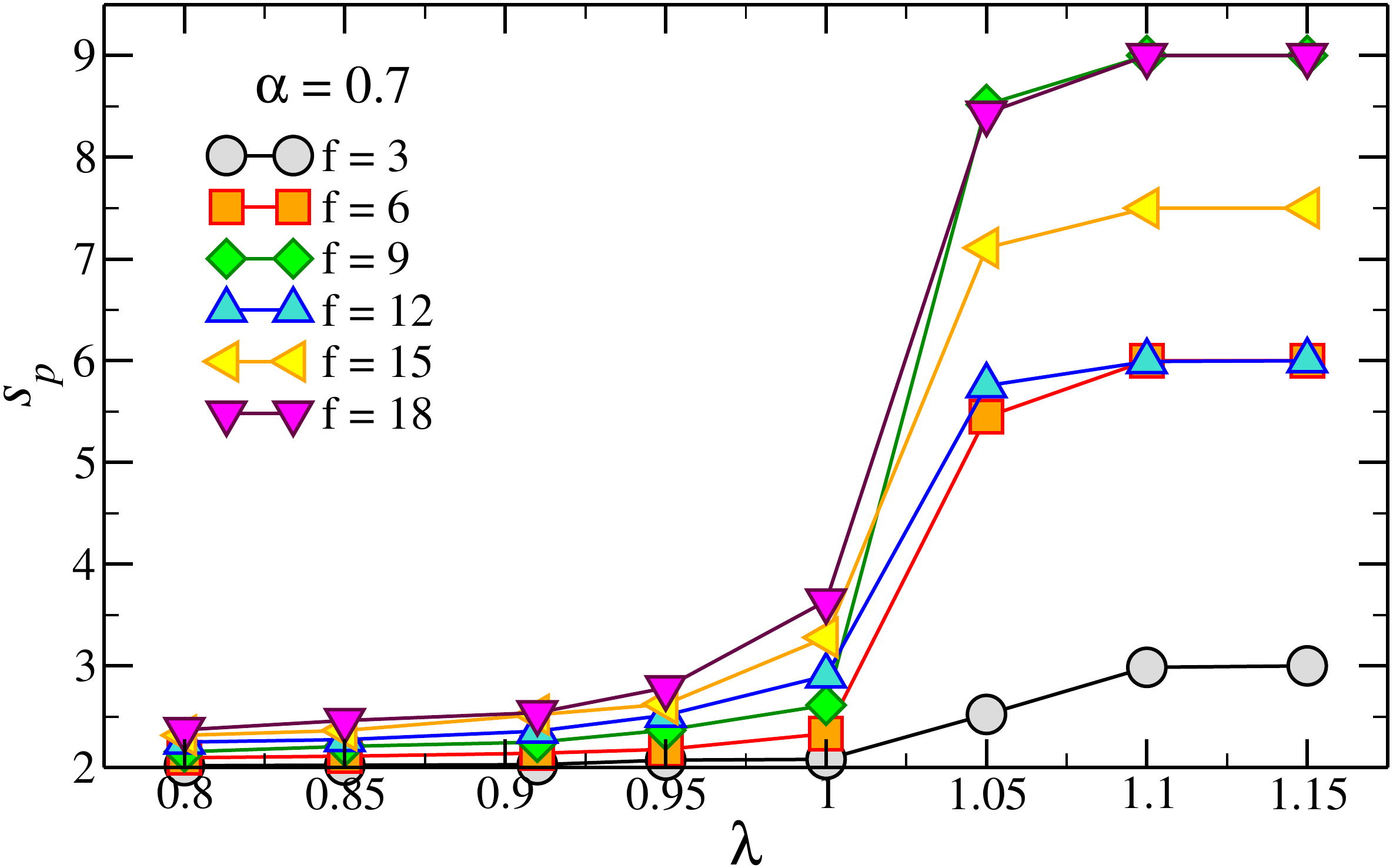}

\caption{\label{fig:size_alpha}Patch population $s_p$ as a function of $\lambda$ for (top) $\alpha=0.5$ and (bottom) 
$\alpha=0.7$.}
\end{figure}

We now move on to the average patch population $s_p$, which is defined as the average number of arms per patch. 
Figure~\ref{fig:size_alpha} shows $s_p$ for $\alpha=0.3$ and $0.5$. At low $\lambda$-values
all curves approach $2$, which is the minimum value according to our definition of a patch. For all the investigated
state points $s_p$ is, within the statistical error, monotonic with $\lambda$. The observed growth of $N_p$ at
intermediate values of $\lambda$ is thus accompanied by an increase of $s_p$, which then plateaus as
the systems undergo a dynamical arrest for $\lambda > 1.15$. At this stage all the arms are involved in a patch, 
thereby $s_p \to f/N_p$.

\begin{figure}[h!]
\includegraphics[width=0.4\textwidth]{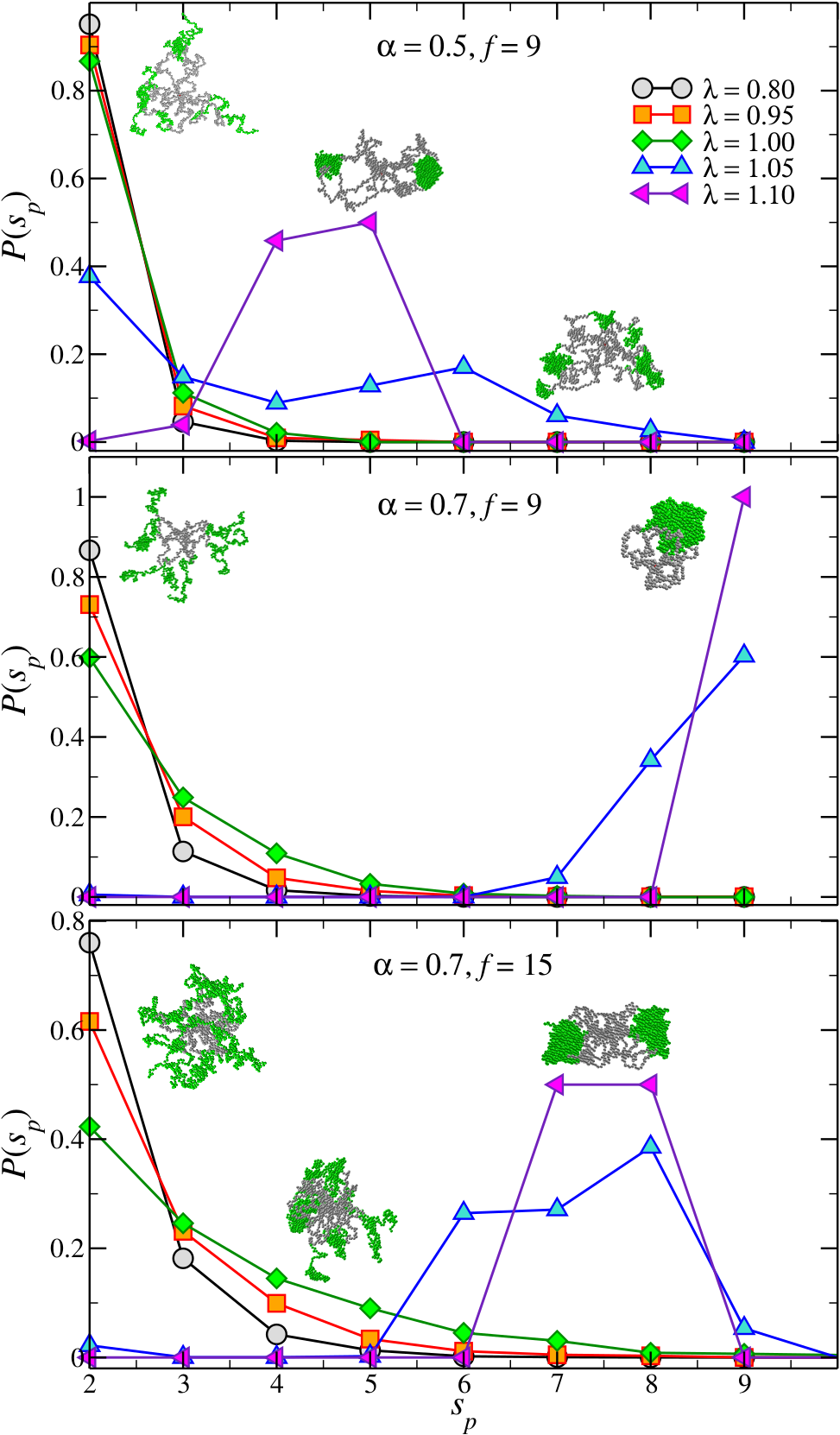}
\caption{\label{fig:sp_distr}Patch population distribution $P(s_p)$ for a TSP with (top) $f=9$, $\alpha=0.5$, 
(middle) $f=9$, $\alpha=0.7$ and (bottom) $f=15$, $\alpha=0.7$. The snapshots show configurations of systems 
with typical values of the patch population.}
\end{figure}

The formation of specific macroscopic phases with the desired properties and symmetry often requires building 
blocks with a well-definite valence and patch size~\cite{bian,flavio_tetra}. We thus have to be sure that all 
the relevant quantities yield not only the right average values but also small fluctuations.
We estimate the conformation fluctuations by looking at the patch population distribution $P(s_p)$, shown in 
Figure~\ref{fig:sp_distr}. At small $\lambda$ 
all the distributions are peaked at $s_p = 2$. As $\lambda$ increases $P(s_p)$ starts developing 
longer and longer tails, rendering the distribution very wide and almost flat. For even stronger attractions, 
$P(s_p)$ becomes non-monotonic and more and more peaked; this non-monotonicity is characteristic of systems 
forming aggregates with a preferential size. Indeed, systems undergoing self-assembly processes, such as 
micelle-formation, have cluster-size distributions which exhibit similar behaviour~\cite{panagiotopoulos}. The 
particular value $\lambda_c$ at which the self-assembly of the patches occurs, i.e. at which most of the arms
are part of a patch, depends on $\alpha$ but not, or 
very weakly, on $f$, and it roughly coincides with the $\lambda$-value at which the number of patches and the
average $s_p$ reach the first plateau in Figs.~\ref{fig:alpha} and~\ref{fig:size_alpha}. Indeed, $\lambda_c$ 
decreases from $\approx 1.15$ for $\alpha=0.3$ to $\approx 1.00$ for $\alpha=0.7$, while its dependence on $f$ 
is negligible.

\begin{figure*}
\begin{center}
\begin{tabular}{cc}
\includegraphics[width=0.4\textwidth]{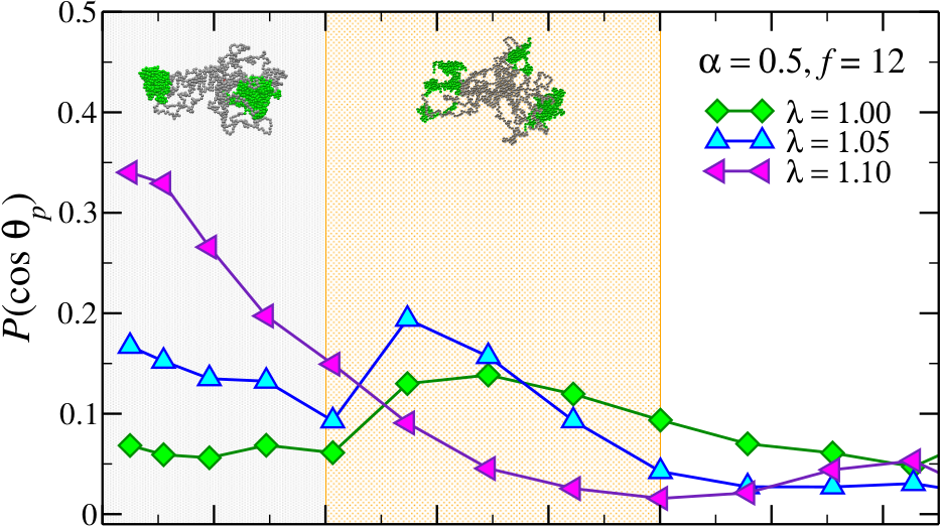} &
\includegraphics[width=0.4\textwidth]{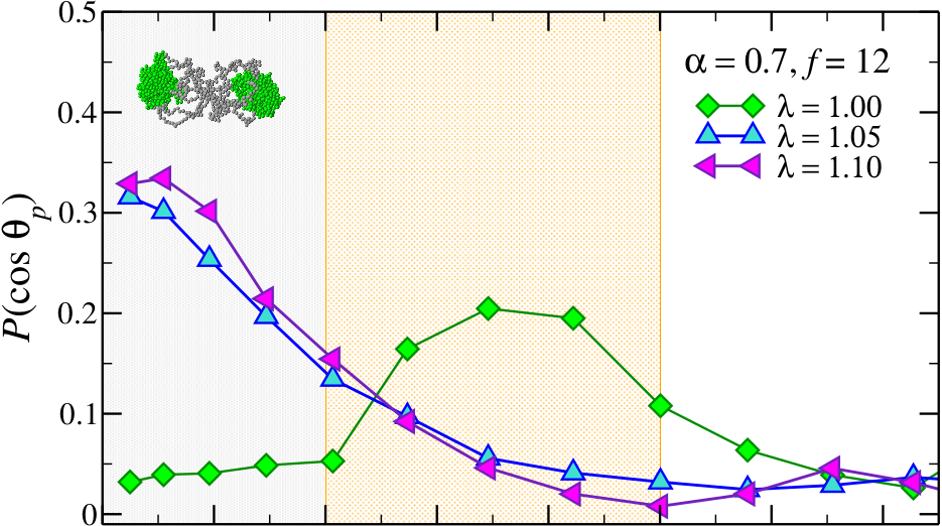}\\
\includegraphics[width=0.4\textwidth]{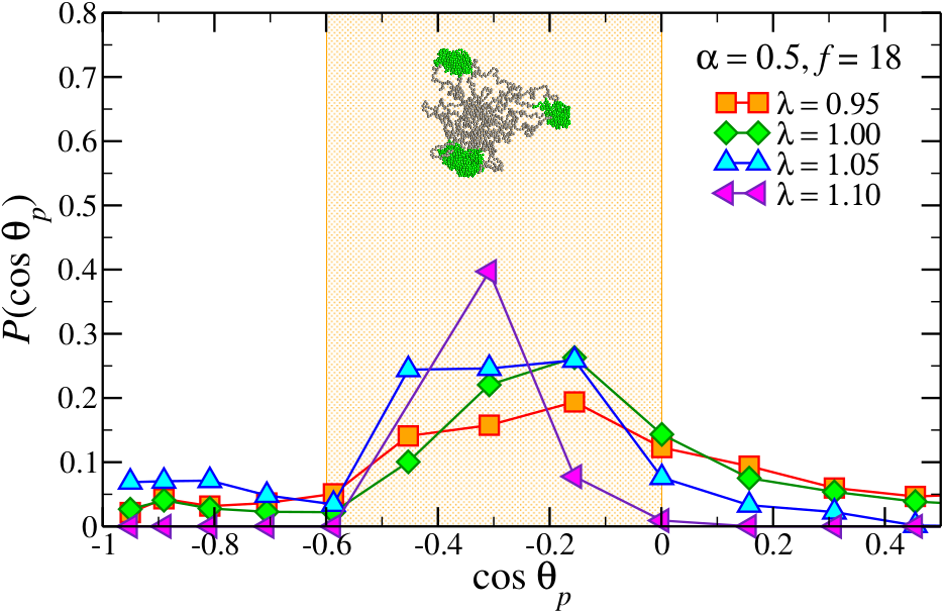} &
\includegraphics[width=0.4\textwidth]{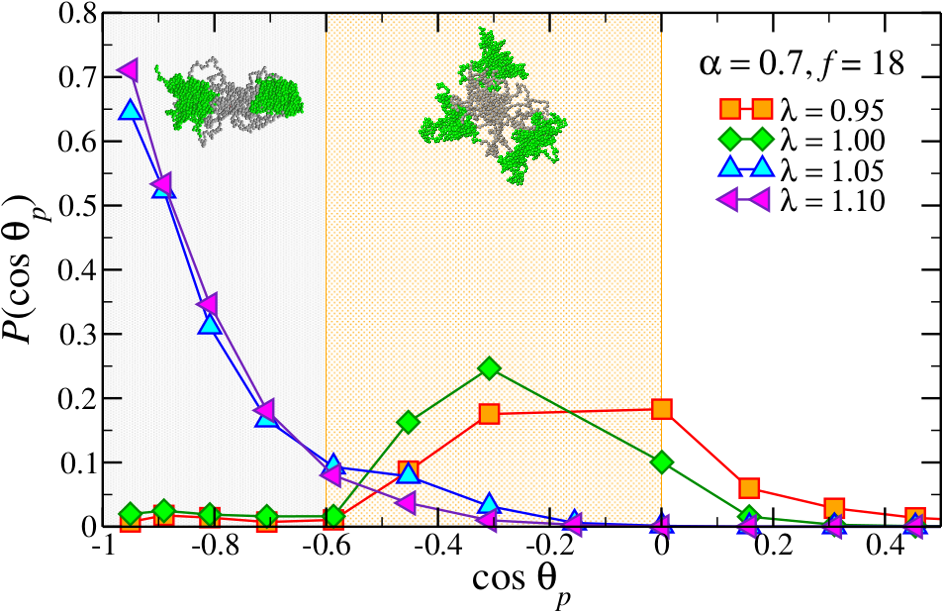}\\
\end{tabular}
\end{center}
\caption{\label{fig:thetap_distr}Distribution of the angle between patches, $P(\theta_p)$, for a TSP with (top) 
$f=12$ and (bottom) $f=18$ for high values of $\lambda$ and for (left) $\alpha=0.5$ and (right) $\alpha=0.7$. 
The snapshots show configurations of systems with typical average angles.}
\end{figure*}

Results obtained with toy models have shown that the phase behaviour of patchy systems, and in particular the 
symmetry of the ordered or partially ordered phases, is determined not only by the number of patches, but 
also by their size and geometrical arrangement~\cite{Rom09a,doppelbauer_patchy_genetic}. 
In addition, internal flexibility has been proven to play a fundamental role in the thermodynamics of
these systems~\cite{frank_nat_phys}. It is thus very important to characterise the patch arrangement. We start off
by introducing a vector $\mathbf{r}_p^i$ that connects the centre of mass of the $i$-th patch to the star centre. We then define
for each pair of patches $i$ and $j$ an angle $\theta_p = \arccos(\mathbf{r}_p^i \cdot \mathbf{r}_p^j)$.
Figure~\ref{fig:thetap_distr} shows the distribution of the cosine of this angle, $P(\cos\theta_p)$, 
for different values of $\alpha$, $f$ and $\lambda$. All the curves are clearly peaked around values 
that directly reflect the number of patches of the nanoparticle: for two and three patches the arrangement 
is planar and hence the average angle is slightly smaller than $\pi$ and $2\pi/3$, respectively. For the 
$f=12$, $\alpha=0.5$ we also observe a coexistence between the two conformations: the number of patches of 
the nanoparticle continuously changes between two and three, giving rise, for intermediate values of $\lambda$, 
to a double-peaked $P(\cos\theta_p)$. For higher values of $\alpha$ or $f$ this transition 
happens in a narrower range of $\lambda$-values and we do not observe any double-peaked distribution for 
the investigated parameters. The effect of the functionality on the distribution of the angle is also 
interesting: as $f$ increases the distributions become more and more peaked, due to the higher local density 
of monomers close to the anchoring point.

\begin{figure}[h!]
\includegraphics[width=0.4\textwidth]{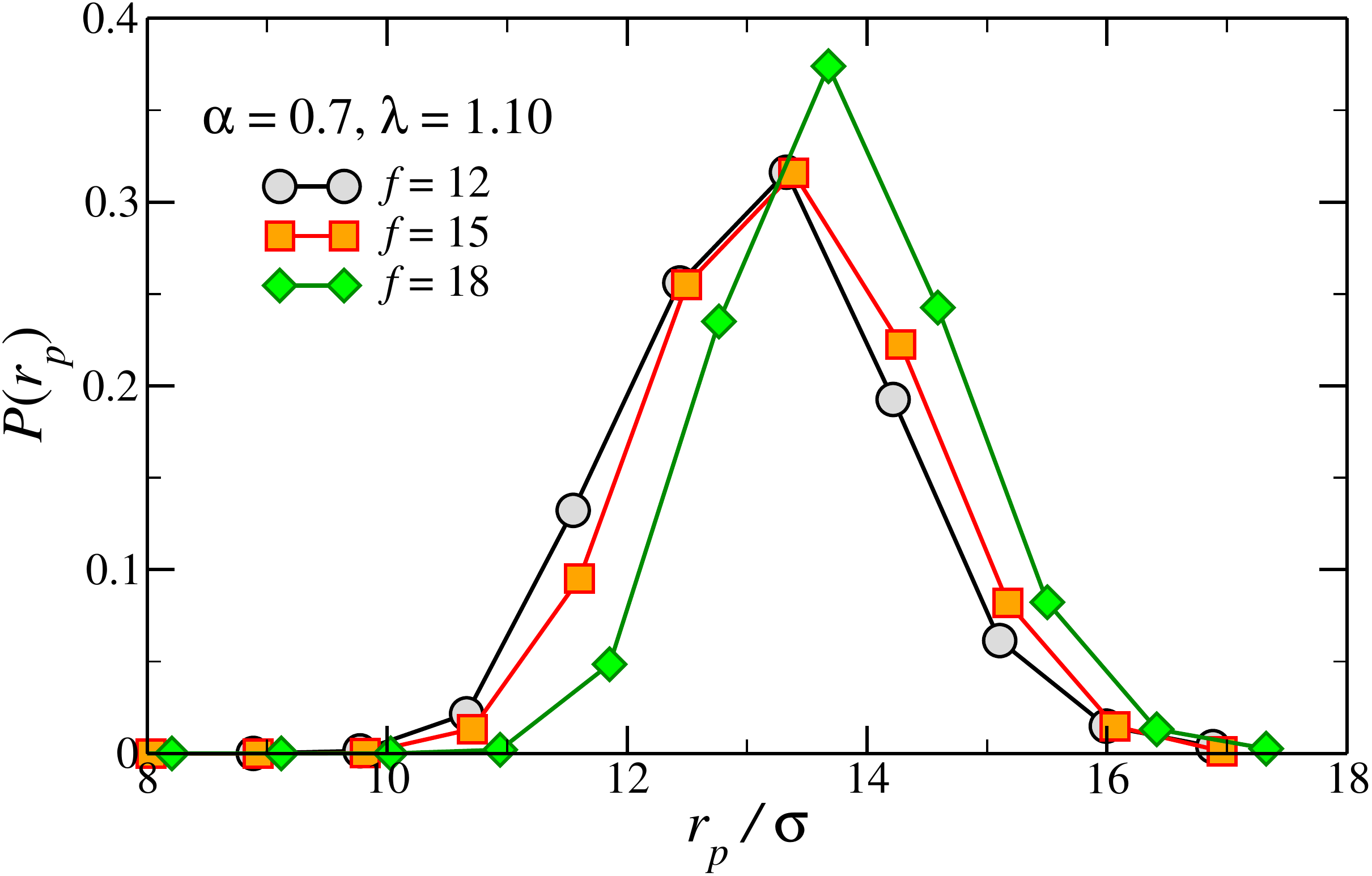}
\caption{\label{fig:rp}Distribution of the distance between a patch and the TSP's anchoring, $P(r_p)$, for 
$f=12$, $15$ and $18$, $\lambda=1.10$ and $\alpha=0.7$.}
\end{figure}

The functionality $f$ plays a similar role in determining the distribution of the radial patch-anchor 
distance $r_p$, $P(r_p)$, which is linked to the stiffness of the particle. Figure~\ref{fig:rp} shows 
$P(r_p)$ for fixed $\alpha=0.7$, $\lambda=1.10$ and three different functionalities, $f=12$, $15$ and 
$18$, chosen so as to yield the same number of patches, $N_p=2$. As $f$ increases we observe a monotonic 
growth of the average patch-anchor distance and a narrowing of the distribution. The net effect of $f$ 
is thus to stiffen and elongate the nanoparticle. 

As a consequence of the above single-particle properties, two networks built with stars with different 
functionalities but same number of patches would differ in the spacing between neighbours and in the
overall stiffness. Indeed, both are increasing functions of $f$, due to the narrowing of the angle and 
radial distributions. On the contrary, the topology, being primarily determined by the number and 
arrangement of the patches, would be less affected by $f$. We thus provide an additional degree of
control on these hierarchically self-assembled materials: the mechanical and elastic properties of
bulk materials can be tuned to a certain degree without varying their topology. In other words, stars 
with different functionalities can be exploited to obtain phases which are similar from the structural
point of view but behave differently, e.g. under shear.

\subsection{Characterisation of the shape}
\label{subsec:shape}

All the results above have been obtained by employing our specific definition of a patch given in
Section~\ref{sec:methods}. Even though the results themselves, as well as visual inspections of the
conformations, confirm that the definition we use is self-consistent, there is always an intrinsic 
ambiguity when dealing with threshold-based cluster algorithms. Therefore, it is important to also 
characterise the star conformation in a way that does not depend upon our specific definition of a 
patch. We do this by computing the gyration tensor and the resulting shape parameters, as defined in 
Eqs.~(\ref{eq:def_delta})-(\ref{eq:def_c}).

\begin{figure}[h!]
\includegraphics[width=0.4\textwidth]{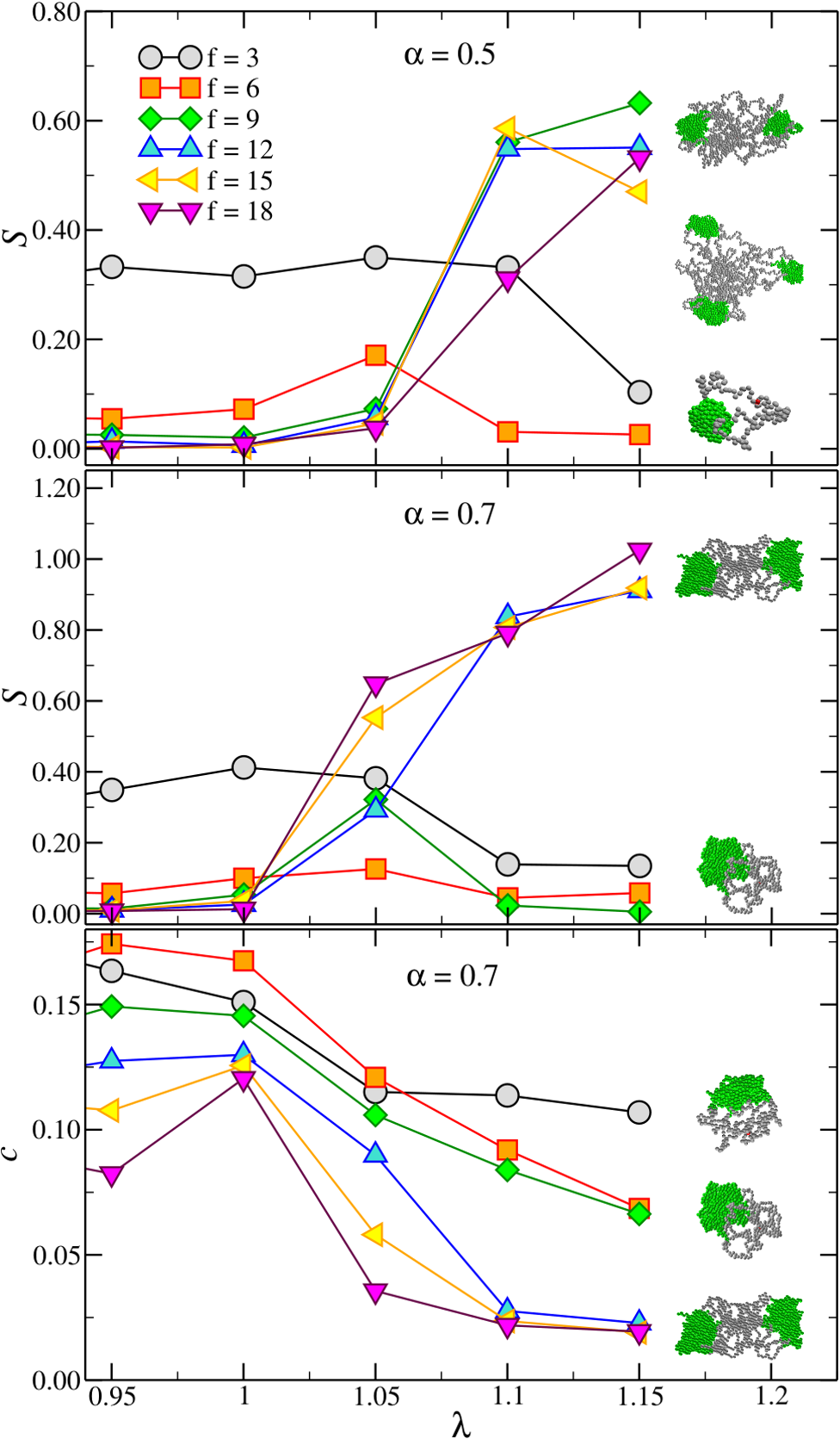}
\caption{\label{fig:shape}Prolateness parameter $S$ for (top) $\alpha=0.5$ and (middle) $\alpha=0.7$ and 
(bottom) acilindricity parameter $c$ for $\alpha=0.7$ for all the investigated functionalities. The snapshots
on the right show typical configurations of high-$\lambda$ systems.}
\end{figure}

Figure~\ref{fig:shape} shows the dependence of the shape parameters on $f$, $\alpha$ and $\lambda$. We start off
by considering the prolateness parameter $S$ which, as noted in Section~\ref{sec:methods}, has the same qualitative
behaviour as the asphericity $\delta$, which is thus not shown here. $S$ is always positive, indicating that 
the stars are always prolate, regardless of $f$, $\alpha$ and $\lambda$, as also observed for chain- and 
ring-polymers~\cite{Narros2013}. Comparing Figure~\ref{fig:shape} with Figure~\ref{fig:size_alpha} 
shows that the steep increase of $S$ at high values of $f$ and $\lambda$ is associated with the presence of two 
patches which, as also shown in Figure~\ref{fig:snapshots}, result in dumbbell-like, very prolate conformations. 
By contrast, stars with smaller functionalities end up in almost spherical single-patch states having $S \approx 0$.
As a consequence, at high values of $\lambda$ there is a clustering of the curves with different $f$, depending 
solely on the number of patches and not on the functionality. We note that the $f=18$, $\alpha=0.7$ case, which form 
$4$ patches at intermediate $\lambda$-values (see Figure~\ref{fig:alpha}), exhibits a very small prolateness, 
demonstrating that $S$ is sensibly different from $0$ only for $N_p \leq 3$. We deduce that the number of patches 
controls the overall shape of the star, while $f$ affects the size, stiffness and deformability of assembled 
nanoparticle, as shown in the previous Section.

The last investigated parameter, the acilindricity $c$, is always small and decreases for large values of $\lambda$. 
This demonstrates that stars are mostly symmetrical around the main axis. In agreement with the trends observed for 
the other shape parameters, this tendency is enhanced when stars assemble into dumbbells due to the presence of two 
large patches. Indeed, at high $f$ and $\lambda$, $c \approx 0$.

\section{Conclusions}

Understanding how to manufacture self-assembling building blocks with  specific softness, functionalisation, shape 
and flexibility by tuning a few microscopical details has an extremely important impact on the material science community 
for a two-fold reason: first of all, it allows to  drive a bottom up self-assembly scheme to engineer new materials 
starting from the microscopic symmetries and properties. Secondly, it makes it possible to add external chemical/physical 
parameters that allow to tune even more the properties of such new materials, without the need to re-formulate their 
molecular structure.   
 
Here we showed that the self-assembly of a very promising class of polymeric building blocks, namely 
telechelic star polymers, can be controlled with great precision by changing the functionality or the 
solvophobic-to-solvophilic ratio, as well as by a careful tuning of the temperature. We have studied how 
the number and size of attractive spots on the surface, herein referred to as patches, vary under changing conditions, 
showing that, at low temperature, these quantities exhibit single-peaked, narrow distributions, thereby providing 
a robust route for the generation of microscopic building blocks with specific, well-defined properties. We have
also studied the flexibility and stiffness of the stars, demonstrating that those depend not only on the number of
patches, but also on the functionality. This opens up the possibility of selecting the elastic properties of the
resulting macroscopic phases without changing their topology and average structure.

Reliable thermosensitive flexible patch formation and tunable dependence on the number of the patches on temperature for 
a given molecular unit is a key ingredient that allows to obtain different mesoscopic self-assembling behaviours from the 
same molecular species. As a consequence, different, possibly ordered, structures, as well as diverse viscoelastic 
properties can be obtained with the same microscopic constituents~\cite{capone:njp:2013}.
The results reported here should be considered together with the notion that the conformation of single stars is 
preserved in low-density bulk phases~\cite{capone:njp:2013}. Indeed, in this case a direct link between the 
conformation of the building blocks and the final structure and phase behaviour of the resulting macroscopic material 
can be established, for example by means of coarser-grained models~\cite{soft_patchy_bianchi}, or theoretical 
treatments~\cite{Rovigatti2014}. The investigation of such \textit{hierarchical self-assembly} processes will 
provide an excellent testing ground for the development of new multiscale methods and also guidance to experiments
for the synthesis of smart materials of the next generation.

\section*{Acknowledgments}
LR acknowledges support from the Austrian Research Fund (FWF) through the Lise-Meitner Fellowship M 1650-N27. 
BC acknowledges support from the \"{O}AW through the APART Fellowship 11723. Computer time at the Vienna Scientific
Cluster (VSC) is gratefully acknowledged.

\bibliography{nanopatchy} 

\end{document}